\documentclass{IEEEtran4PSCC}
\ifCLASSINFOpdf
  \usepackage[pdftex]{graphicx}
\else
\fi
\usepackage[cmex10]{amsmath}
\makeatletter
\let\old@ps@headings\ps@headings
\let\old@ps@IEEEtitlepagestyle\ps@IEEEtitlepagestyle
\def\psccfooter#1{%
    \def\ps@headings{%
        \old@ps@headings%
        \def\@oddfoot{\strut\hfill#1\hfill\strut}%
        \def\@evenfoot{\strut\hfill#1\hfill\strut}%
    }%
    \def\ps@IEEEtitlepagestyle{%
        \old@ps@IEEEtitlepagestyle%
        \def\@oddfoot{\strut\hfill#1\hfill\strut}%
        \def\@evenfoot{\strut\hfill#1\hfill\strut}%
    }%
    \ps@headings%
}
\makeatother

\usepackage[T1]{fontenc}
\usepackage[utf8]{inputenc}
\usepackage[hyperindex=true, %
  pdftitle={Experimental Validation of Fully Distributed Peer-to-Peer Optimal Voltage Control with Minimal Model Requirements},%
  pdfauthor={Lukas Ortmann, Alexander Prostejovsky, Kai Heussen, Saverio Bolognani},%
  colorlinks=true,%
  pagebackref=false,%
  plainpages=false,%
  pdfpagelabels,%
  linkcolor=black,%
  citecolor=black,%
  filecolor=black,%
  urlcolor=black%
]{hyperref} 

\usepackage[dvipsnames]{xcolor}
\usepackage{tikz}
\usepackage{pgfplots} 
\usepackage{pgfgantt}
\usepackage{pdflscape}
\pgfplotsset{compat=newest} 
\pgfplotsset{plot coordinates/math parser=false}
\usepackage{wrapfig}
\usepackage{multirow}
\usepackage{caption}
\usepackage{amssymb}
\usepackage{cite}
\usepackage{booktabs}
\usepackage{algorithm}
\usepackage[noend]{algpseudocode}
\usepackage{float}
\usepackage[absolute,overlay]{textpos}

\usepackage[font=small]{caption}

\newlength\fwidth
\newlength\fheight
\newlength\fmargin
\usepackage[prependcaption,colorinlistoftodos]{todonotes}

\definecolor{highlight}{RGB}{0,0,0}

\begin{document}
%
\title{Experimental Validation of Fully Distributed Peer-to-Peer Optimal Voltage Control with Minimal Model Requirements}

\author{
\IEEEauthorblockN{Lukas Ortmann\\
Saverio Bolognani}
\IEEEauthorblockA{Automatic Control Laboratory \\
ETH Zurich\\
Zurich, Switzerland
}
\and
\IEEEauthorblockN{Alexander Prostejovsky\\ 
Kai Heussen}
\IEEEauthorblockA{Department of Electrical Engineering \\
Technical University of Denmark (DTU)\\
Roskilde, Denmark
}
}

\maketitle

\begin{textblock*}{\textwidth}(17mm,5mm) 
	\centering \bf \textcolor{NavyBlue}{Published as "Fully Distributed Peer-to-Peer Optimal Voltage Control with Minimal Model Requirements" on \emph{Electric Power Systems Research,} vol. 189, December 2021.\\\url{https://doi.org/10.1016/j.epsr.2020.106717}}
\end{textblock*}

\begin{abstract}
\color{highlight}
This paper addresses the problem of voltage regulation in a power distribution grid using the reactive power injections of grid-connected power inverters.
We first discuss how purely local voltage control schemes cannot regulate the voltages within a desired range under all circumstances and may even yield detrimental control decisions. Communication and, through that, coordination are therefore needed. 
On the other hand, short-range peer-to-peer communication and knowledge of electric distances between neighbouring controllers are sufficient for this task.
We implement such a peer-to-peer controller and test it on a 400~V distribution feeder with asynchronous communication channels, confirming its viability on real-life systems.
Finally, we analyze the scalability of this approach with respect to the number of agents on the feeder that participate in the voltage regulation task.
\end{abstract}


\thanksto{
\setlength{\columnsep}{6pt}%
\setlength{\intextsep}{3pt}%
\begin{wrapfigure}{o}{0.14\columnwidth}
  \raggedleft
  \vspace{-4pt}
    \includegraphics[width=0.14\columnwidth]{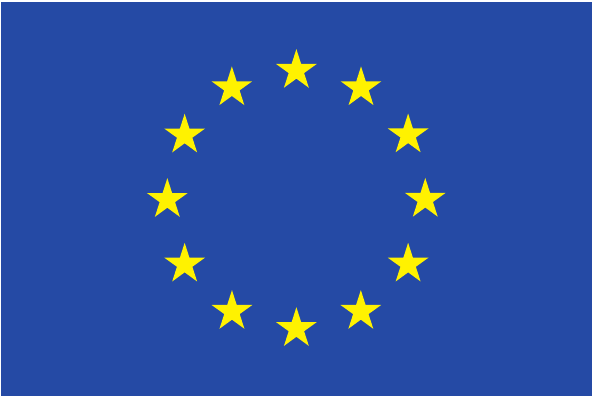}
\end{wrapfigure}
This research has been performed using the ERIGrid Research Infrastructure and is part of a project that has received funding from the European Union's Horizon 2020 Research and Innovation Program under the Grant Agreement No. 654113. The support of the European Research Infrastructure ERIGrid and its partner Technical University of Denmark is very much appreciated.
This paper reflects only the authors' view and the EU Commission is not responsible for any use that may be made of the information it contains.\\
\hspace*{\parindent} The research leading to this work was supported in part by the Swiss Federal Office of Energy grant \#SI/501708 UNICORN.\\
\hspace*{\parindent} Corresponding author: Lukas Ortmann, email: \href{mailto:ortmannl@ethz.ch}{ortmannl@ethz.ch}.
}

\section{Introduction}

Future power distribution grids are expected to host a significant portion of the total generation capacity, for the most part from renewable energy sources like solar and micro-wind installations.
Meanwhile, the deployment of a distributed electric mobility infrastructure will substantially increase the loading of this infrastructure.
This transition will inevitably affect the operating regime of distribution feeders, and will increase the risk of both overvoltage and undervoltage contingencies.
On the other hand, microgenerators and electric vehicle charging stations will offer unprecedented voltage control flexibility via their power inverters, offering a finely distributed network of reactive power compensators.

\begin{figure}[b]
    \centering
    \includegraphics[scale=0.8]{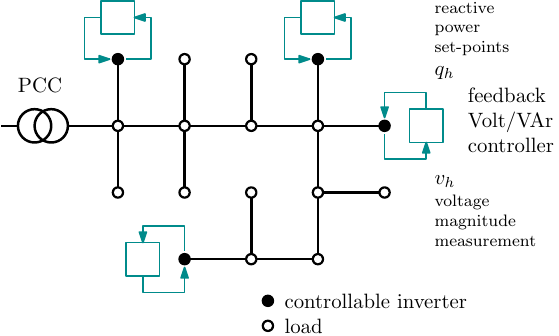}
    \caption{Schematic representation of the control architecture employed by fully decentralized Volt/VAr feedback strategies, e.g. \cite{IEEE1547, VDE, ENTSOE}.}
    \label{fig:decentralized}
\end{figure}

For the control of these reactive power compensators, a multitude of decentralized Volt/VAr feedback control strategies have been proposed (e.g., Volt/VAr droop control; cf. \cite{Bolognani2019} for a literature review) and ultimately incorporated in many grid codes and standards \cite{IEEE1547,VDE,ENTSOE}. 
These strategies rely on the control architecture schematically represented in Figure~\ref{fig:decentralized}, where each power inverter independently regulates its reactive power injection based on the voltage measurement performed at its point of connection, typically via a static update map
\[
q_h(t+1) = f_h (v_h(t)).
\]
The update map $f_h$ is usually the outcome of heuristic design procedures. In most cases the design is completely model-free (no grid information is used), although computational design approaches have also been proposed \cite{Karagiannopoulos2019TSG}.

Fully decentralized feedback control solutions present multiple advantages, such as:
\begin{itemize}
\item high robustness, given by the absence of a single point of failure;
\item economical deployment and retrofitting (plug-and-play);
\item minimal actuation time delays, due to the absence of any communication; 
\item modularity and interoperability, as individual inverters do not coordinate their action;
\item scalability and computational simplicity.
\end{itemize}

However, purely decentralized control strategies fail to ensure feasible voltages, even if such a feasible solution exists, as recently proven in \cite{Bolognani2019}.
Conversely, centralized feedback Volt/VAr solutions are guaranteed to drive the system to a feasible voltage profile, using the same measurements collected in the decentralized setting (i.e., only voltage magnitude measurements of the inverters) but processing them in a centralized manner.
We refer to \cite{Doerfler2019,Bernstein2019} for a recent review of \emph{feedback optimization} methods that can be employed to design these centralized feedback Volt/VAr strategies, and to \cite{Ortmann2020} for an experimental validation that demonstrates a remarkable robustness against model uncertainty.
The disadvantages of centralized feedback optimization are that a communication channel between a central computational unit and all the power inverters is required, and a global model of the grid needs to be known at this central location.

This paper is motivated by a fundamental question: \textcolor{highlight}{is it possible to achieve optimal Volt/VAr regulation without collecting all measurement and all model information in a centralized location?}
A limited number of recent works contributed towards an answer to this question by proposing feedback control strategies that are extremely parsimonious in terms of information that inverters need to communicate:
\begin{itemize}
    \item A distributed solution for the voltage regulation and loss minimization problem is proposed in \cite{Bolognani2015}, allowing asynchronous communication between agents (but relying on both angle and magnitude measurements).
    \item In \cite{vancutsem2016}, power inverters are controlled by individual automata that communicate a ``distress signal'' only when their regulation problem becomes infeasible; however, this strategy is not guaranteed to converge to the optimal regulation.
    \item In a similar spirit, \cite{Magnusson2019triggered} proposes a distributed strategy in which inverters communicate only when triggered by local voltage violation rules; an all-to-all communication channel is however assumed.
    \item A primal-dual method that requires only communication between neighboring inverters is proposed in \cite{Qu2019};
    \item The authors of \cite{Magnusson2019} demonstrate how coordination between inverters can be achieved by only transmitting a few bits of information;
    \item a distributed dual ascent method is employed in \cite{magnusson2019distributed}, allowing for delayed communication between inverters;
    \item finally, \cite{Bolognani2019} proposes a distributed synchronous dual ascent method with a nested quadratic program.
\end{itemize}

To the best of the authors' knowledge, none of these distributed solutions has been implemented and tested on a real grid with physically distributed computations. 

In this work, we provide a proof-of-concept demonstration of how Volt/VAr regulation can be achieved via a distributed feedback control law, namely under the specifications that:
\begin{itemize}
    \item each inverter can only establish \textcolor{highlight}{asynchronous} peer-to-peer communication with its neighboring inverters;
    \item each inverter only maintains model information regarding its grid neighborhood;
    \item no central coordination unit is present.
\end{itemize}

The reported experiment also validates other important features of this distributed solution such as its robustness against noisy measurements, its real-time computational feasibility, and the viability of algorithm distribution in a peer-to-peer setting with no master algorithm synchronization. 

\textcolor{highlight}{Finally, we investigated scalability of the proposed approach via a series of numerical experiments.}

\section{Distributed Voltage Control}

In this section we report the procedure proposed in \cite{Bolognani2019} to design a distributed controller for the Volt/VAr regulation problem. 
\textcolor{highlight}{Although a synchronous communication channel was assumed in \cite{Bolognani2019}, it provides the key idea on how to achieve optimal coordination via only short-range exchange of information.}

\subsection{Feedback Optimization Controller}

The controller is derived from the optimization problem
\begin{equation}
\begin{alignedat}{2}
\min &\quad\frac{1}{2}q^TMq\\
\text{subject to} &\quad v_\textrm{min}\le v_h(q,w)\le v_\textrm{max} \quad &&\forall h\\
&\quad q_\textrm{min}\le q_h\le q_\textrm{max} && \forall h.
\end{alignedat}
\label{eq:optimization_problem}
\end{equation}
where the matrix $M$ is a square, symmetric and positive definite design parameter and $v$ and $q$ are the vectors we obtain by stacking the voltages $v_h$ and reactive power set-points $q_h$ of the different inverters, respectively. The function $v_h(q,w)$ is the steady-state map of the nonlinear power flow equations that defines voltages $v_h$ as a function of both reactive powers $q$ and external influences $w$ (e.g., active and reactive demands, active generation).

{\color{highlight}
Active power injections are not a decision variable in \eqref{eq:optimization_problem} for the following reason.  Controlling the active power of devices comes with an  economic cost, whereas the usage of reactive power is free (neglecting the active power losses generated by the reactive power flows). Therefore, it is typically preferred to use the reactive power capabilities in the network to their full extent before controlling active power injections.}

To solve \eqref{eq:optimization_problem} we introduce the dual multipliers $\lambda_{h,\textrm{min}}$ and $\lambda_{h,\textrm{max}}$ for the voltage constraints of every inverter $h$. Stacking them gives us the vector $\lambda = \left[\begin{smallmatrix} \lambda_\textrm{min}\\\lambda_\textrm{max} \end{smallmatrix}\right]$ 
with which we form the Lagrangian $\mathbf{L}(q,\lambda)$ by dualizing the voltage constraints:
\begin{align}
\begin{split}
\mathbf{L}(q,\lambda)=\frac{1}{2}q^TMq&+\sum_{h}\lambda_{h,\textrm{min}}(v_\mathrm{min}-v_h(q,w))\\&+\sum_{h}\lambda_{h,\textrm{max}}(v_h(q,w)-v_\mathrm{max}).
\end{split}
\end{align}
We thus define the equivalent dual optimization problem
\begin{align}
\begin{split}
\max_{\lambda \ge 0} \; \min_q&\quad\mathbf{L}(q,\lambda)\\
\text{subject to}& \quad q_{h,\text{min}} \le q_h \le q_{h,\text{max}} \quad \forall h.
\end{split}
\label{eq:dual_problem}
\end{align}
The optimization problems \eqref{eq:optimization_problem} and \eqref{eq:dual_problem} have the same solution (Strong Duality Theorem, \cite[Proposition 5.3.2]{Bertsekas1999}).
We adopt an iterative dual ascent update on $\lambda$ to compute the solution of \eqref{eq:dual_problem}, obtaining
\begin{align}
\begin{split}
\lambda_\text{min}(t+1) &= [\lambda_\text{min}(t)+\alpha\nabla_{\lambda_\text{min}}\mathbf{L}(q(t),\lambda(t))]_{\geq 0}\\
&= [\lambda_\text{min}(t)+\alpha(v_\text{min}-v(q(t),w))]_{\geq 0}\\
\lambda_\text{max}(t+1)&=[\lambda_\text{max}(t)+\alpha\nabla_{\lambda_\text{max}}\mathbf{L}(q(t),\lambda(t))]_{\geq 0}\\
&=[\lambda_\text{max}(t)+\alpha(v(q(t),w)-v_\text{max})]_{\geq 0}.
\end{split}
\label{eq:lambda}
\end{align}
As we can see every inverter integrates its own voltage violation with a gain of $\alpha$.
This corresponds to the integral part of a PI-controller and can be done locally, by using feedback from the physical system through voltage magnitude measurements $v(t)$ of the inverters, rather then via a numerical evaluation of $v(q(t),w)$.
To find the optimal reactive power set-points we use the newly calculated $\lambda(t+1)$ and solve
\begin{equation}
\begin{split}
    q(t+1)=\quad \arg\min_{q} &\quad \mathbf{L}(q,\lambda(t+1)) \\
    \text{subject to}& \quad q_{h,\text{min}} \le q_h \le q_{h,\text{max}} \quad \forall h. 
\end{split}
\label{eq:min_L}
\end{equation}
Towards this goal, we introduce the approximation
\begin{equation} \label{eq:linear_model}
    \frac{\partial v(q,w)}{\partial q} \approx X
\end{equation}
where $X$ is the reduced bus reactance matrix that can be derived from the grid topology and the cable data. The sensitivity described by $X$ is similar to power transfer distribution factors for active power generation on the transmission level.
Under no-load conditions and the assumption of negligible cable resistances this approximation is accurate, because the nonlinearity of the power flow equations is mild near this operating point \cite{Bolognani2015fast}.
In our application the system can be heavily loaded and the cable resistances are high. It was shown in \cite{Ortmann2020} that feedback optimization is sufficiently robust against this model mismatch.

This approximation makes $v(q,w)$ linearly dependent on $q$, and we can approximate \eqref{eq:min_L} with a convex quadratic optimization problem (QP).
This QP involves the decision variables of all DERs and can be solved by collecting all the necessary information (the multipliers $\lambda(t+1)$ and the parameters $X$) in a central control unit \cite{Ortmann2020}.
In the following we use the idea proposed in \cite{Bolognani2019} to show how \eqref{eq:min_L} can also be solved in a distributed manner, without centralized computation or centralized model knowledge.

\subsection{Distributing the Controller}
\label{ssec:distributing}

To solve the subproblem \eqref{eq:min_L} in a distributed manner we perform $K$ iterative steps, which will have to be executed between the times $t$ and $t+1$. To denote these iterative steps we introduce a new iteration counter $\tau$. We also introduce the dual multipliers $\mu_{h,\textrm{min}}$ and $\mu_{h,\textrm{max}}$ for the reactive power constraints of every inverter $h$, which we stack in the vector
$\mu = \left[\begin{smallmatrix} \mu_\text{min}\\\mu_\text{max} \end{smallmatrix}\right]$.
By dualizing the reactive power constraints, we define the Langrangian 
\begin{align}
\begin{split}
    \mathbf{L}_\lambda(q,\mu)=\mathbf{L}(q,\lambda) &+\sum_{h}\mu_{h,\textrm{min}}(q_\mathrm{min}-q_h)\\&+\sum_{h}\mu_{h,\textrm{max}}(q_h-q_\mathrm{max})
\end{split}
\end{align}
and the following optimization problem:
\begin{equation}
	\max_{\mu \ge 0} \; \min_q  \ \  \mathbf{L}_\lambda(q,\mu).
	\label{eq:dualoptimizationmuq}
\end{equation}
The optimization problems \eqref{eq:min_L} and \eqref{eq:dualoptimizationmuq} have the same solution (Strong Duality Theorem, \cite[Proposition 5.3.2]{Bertsekas1999}).
Similarly as before, we solve this optimization problem via  gradient ascent iterations on $\mu$ with step size $\gamma$:
\begin{align}
\begin{split}
\mu_\text{min}(\tau+1)
&=[\mu_\text{min}(\tau)+\gamma\nabla_{\mu_\text{min}}\mathbf{L}_\lambda(\hat q(\tau),\mu(\tau))]_{\geq 0}\\
&=[\mu_\text{min}(\tau)+\gamma(q_\textrm{min}-\hat q(\tau))]_{\geq 0}\\
\mu_\text{max}(\tau+1)&=[\mu_\text{max}(\tau)+\gamma\nabla_{\mu_\text{max}}\mathbf{L}_\lambda(\hat q(\tau),\mu(\tau))]_{\geq 0}\\
&=[\mu_\text{max}(\tau)+\gamma(\hat q(\tau)-q_\textrm{max})]_{\geq 0}
\end{split}
\label{eq:mu_update}
\end{align}
where
\[
\hat q(\tau)
= \arg\min_q  \mathbf{L}_\lambda(q,\mu(\tau)).
\]
Observe, that the update of $\mu_\textrm{min}$ and $\mu_\textrm{max}$ can be done locally by every inverter by integrating the constraint violation of the virtual quantity $\hat q(\tau)$.
In order to compute the unconstrained minimizer $\hat q(\tau)$, we take the derivative $\nabla_q \mathbf{L}_\lambda(q,\mu)$ and obtain
\begin{equation}
\begin{split}
\nabla_q \mathbf{L}_\lambda(q,\mu) 
 = Mq +\frac{\partial v}{\partial q} (\lambda_\text{max} - \lambda_\text{min}) 
 + \mu_{\text{max}}-\mu_{\text{min}}.
\end{split}
\label{eq:derivative_Lagrangian}
\end{equation}
As stated before, we approximate the derivative $\partial v / \partial q$ with $X$ and set \eqref{eq:derivative_Lagrangian} to 0. We then solve for $q$ and obtain
\begin{equation}
    \begin{split}
        \hat q(\tau) &= -M^{-1}X(\lambda_\text{max}-\lambda_\text{min}) \\
        &+M^{-1}[\mu_{\text{max}}(\tau)-\mu_{\text{min}}(\tau)].    
    \end{split}
    \label{eq:qhat}
\end{equation}
Equation \eqref{eq:qhat} reveals that all the communication requirements of the proposed iterative algorithm are encoded in the sparsity of the matrices $M^{-1}$ and $M^{-1}X$.
In fact, off-diagonal non-zero elements of these two matrices determine components of $\lambda$ and $\mu$ that need to be communicated between inverters in order to compute $\hat q(\tau)$.

\begin{figure}[tb]
    \centering
    \includegraphics[scale=0.8]{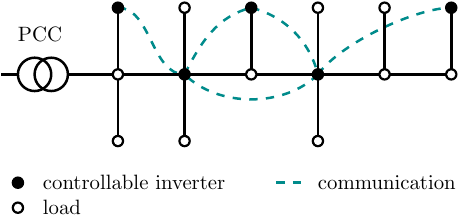}\\[2mm]
    \caption{Schematic representation of neighbor-to-neighbor communication, where we adopt the definition of neighbors from \cite{Bolognani2015}:
    two inverters are neighbors if the electrical path connecting them does not pass through any other bus where a controlled inverter is connected.}
    \label{fig:neighbors}
\end{figure}

In order to maximize the sparsity of both these matrices, we exploit the structure inherited from the physical system. 
\textcolor{highlight}{We inherit the formal definition of neighboring inverters from \cite{Bolognani2015}, see Figure~\ref{fig:neighbors}. 
Neighbors according to this definition can be conveniently discovered via correlation analysis of the voltage measurements, even without central supervision (see \cite{Bolognani18,DEKA2018} and references therein). 
}

Due to the sparsity of the power flow equations, $G=X^{-1}$ is a sparse matrix: namely, $G_{ij}$ of $G$ is non-zero only if the buses $i$ and $j$ are neighbors, and $G_{ij}$ depends only on the electrical impedance of the path between $i$ and $j$.

Based on this observation, we choose $M=X$ which yields
\begin{equation}
\hat q(\tau) = \lambda_\text{min} - \lambda_\text{max}
+G\left( \mu_{\text{min}}(\tau)-\mu_{\text{max}}(\tau)\right).
\label{eq:q_update}
\end{equation}
Therefore inverter $i$ only needs to gather  $\mu_{j,\text{min}}$ and $\mu_{j,\text{max}}$ from their neighbors $j$ to calculate $\hat q_i(\tau)$.
\textcolor{highlight}{
Note that $M=X$ is possibly not the only choice that allows to distribute the algorithm, if one accepts to use a descent direction in the gradient steps which is not the steepest one \cite[Section III.F]{Doerfler2019}.
}
\textcolor{highlight}{
With a proper choice of the gain $\gamma$ (for which we refer to Section~\ref{sec:scaling}) the alternate execution of \eqref{eq:mu_update} and \eqref{eq:q_update} is guaranteed to converge to the solution to \eqref{eq:dualoptimizationmuq}.
We assume that the number of iterations $K$ is chosen sufficiently large so that, after $K$ iterations, $\hat q$ is accepted as the solution to \eqref{eq:min_L} and determines the next set-point $q(t+1)$. The effect of this approximation is also studied in Section~\ref{sec:scaling}.}

\begin{figure}[t]
	\centering
	\includegraphics[width=0.95\columnwidth]{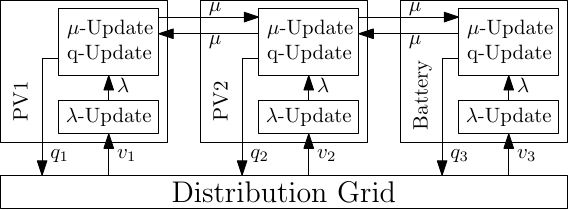}\\[1mm]
	\caption{Control architecture. Measurement and actuation is performed locally by each controller. Only the dual multipliers $\mu_\textrm{min}$ and $\mu_\textrm{max}$ need to be communicated to neighboring peers.}
	\label{fig:block_diagram}
\end{figure}

The resulting control algorithm consists in a main loop, reported hereafter as Algorithm~1, and a nested iterative procedure, Algorithm~2. 
Communication between agents only happens as part of Algorithm~2, when the dual multipliers $\mu$ of the reactive power constraints need to be communicated with neighbors (steps 7--8). All other steps are basic numerical operations that each inverter performs locally. 
The resulting control architecture is represented in Figure~\ref{fig:block_diagram}.

\textcolor{highlight}{Note that the implementation of our controller inherits the theoretical guarantees provided in \cite{Bolognani2019}, including Proposition~6 that guarantees asymptotic optimality (under the linearity condition \eqref{eq:linear_model} and assuming that \eqref{eq:min_L} is solved exactly).}

\begin{algorithm}[H]
\caption{Feedback optimization controller}
\begin{algorithmic}[1]
    \State Initialize: $\lambda_{h,\text{min}}$ and $\lambda_{h,\text{max}}$ with 0
    \Loop
    \State Locally measure the voltage magnitude $v_h$
    \State Locally update $\lambda_{h,\text{min}}$ and $\lambda_{h,\text{max}}$ via \eqref{eq:lambda}
    \State Jointly compute $q_h$ via Algorithm~2
    \State Locally apply the new set-point $q_h$
    \State Wait until next system interrupt\\
    \Comment{System interrupts generated every $T$ seconds}
    \EndLoop
\end{algorithmic}
\end{algorithm}

\begin{algorithm}[H]
\caption{Distributed QP solver}
\begin{algorithmic}[1]
\If{Algorithm~2 was never executed previously}
\State Initialize: $\mu_{h,\textrm{min}}$, $\mu_{h,\textrm{max}}$ and $\hat q_h$ with 0
\Else
\State Keep previous values to warm start
\EndIf
\State counter = 0
\Comment{Iteration counter}
\Repeat
    \State Locally update $\mu_{h,\textrm{min}}$ and $\mu_{h,\textrm{max}}$ via \eqref{eq:mu_update}
    \State Send $\mu_{h,\textrm{min}}$ and $\mu_{h,\textrm{max}}$ to neighbors
    \State Receive $\mu_{i,\textrm{min}}$ and $\mu_{i,\textrm{max}}$ from all neighbors
    \State Locally, compute $\hat q_h$ via \eqref{eq:q_update}
    \State counter = counter + 1
    \Until{counter == K}
\State Return the solution $q_h=\hat q_h$
\end{algorithmic}
\end{algorithm}

\section{Experimental Setup}

The experiment has been implemented in the SYSLAB facility located on the Ris\o{} campus of the Technical University of Denmark.
The setup consists of a $400~V$ three-phase electric grid connecting a variety of DERs (solar panels, wind turbines, a flow battery, a diesel generator, controllable loads, among others).
Each device has an associated computer node running a distributed monitoring and control platform.

\subsection{Algorithm Implementation and Deployment}
\label{ssec:algorithmimplementation}
\textcolor{highlight}{An existing distributed optimization framework developed at DTU \cite{semitsogloutsiapos2019} was adopted to implement the proposed distributed optimization controllers over an asynchronous communication channel.}
Each computer node implements Algorithm~1 in major fixed time intervals of $T=10$~seconds, based on their individual clock. 
This is therefore the rate at which measurements are collected (line 3) and the system is actuated (line 6). 
\textcolor{highlight}{
The choice of such a long interval is due to hardware constraints given by the laboratory setup. A more frequent actuation is often possible. However, the actuation interval should be long enough for the system to settle and reach its steady state. The frequency at which the system can be actuated will always be significantly lower than the rate at which inverters can communicate (see Section~\ref{sec:scaling} for a discussion on the implications on the algorithm scalability).
}

Algorithm~2 is executed in K iterations. 
Lines 7 and 8 of this algorithm require communication between neighbours, where the communication time is variable, dependent on uncontrollable influences. 
Coherency of the algorithm, and thereby a synchronous advancement of the algorithm steps, is achieved by letting individual nodes remain idle  until data has been received from all neighbours (line~8 of Algorithm~2).
This way, the synchronous Algorithm~2 is transparently implemented on an asynchronous communication channel, which has better scaling properties than a synchronous one in such a setup \cite{semitsogloutsiapos2019}.
ZeroMQ \cite{Hintjens2013} is used as the underlying messaging library with TCP transport, facilitating reliable data delivery.
The code comprising the distributed framework and algorithm is deployed to each of the active SYSLAB node computers and operates as a local process. 

\subsection{Test Case and Experiment Design}
\label{ssec:benchmark}

\begin{figure}[tb]
    \centering
    \includegraphics[scale=0.85]{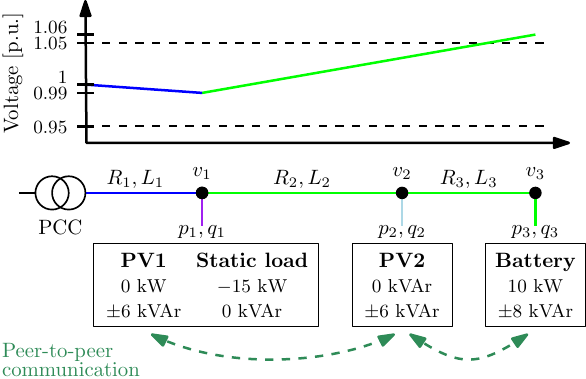}\\[2mm]
    \caption{Sketch of the voltage profile, the distribution feeder and the peer-to-peer communication. The colors of the voltage profile and the diagram match the colors in the topology in Figure~\ref{fig:syslab_topology}.}
    \label{fig:voltage_profile}
\end{figure}

\begin{figure}[tb]
\centering
\includegraphics[width=\columnwidth]{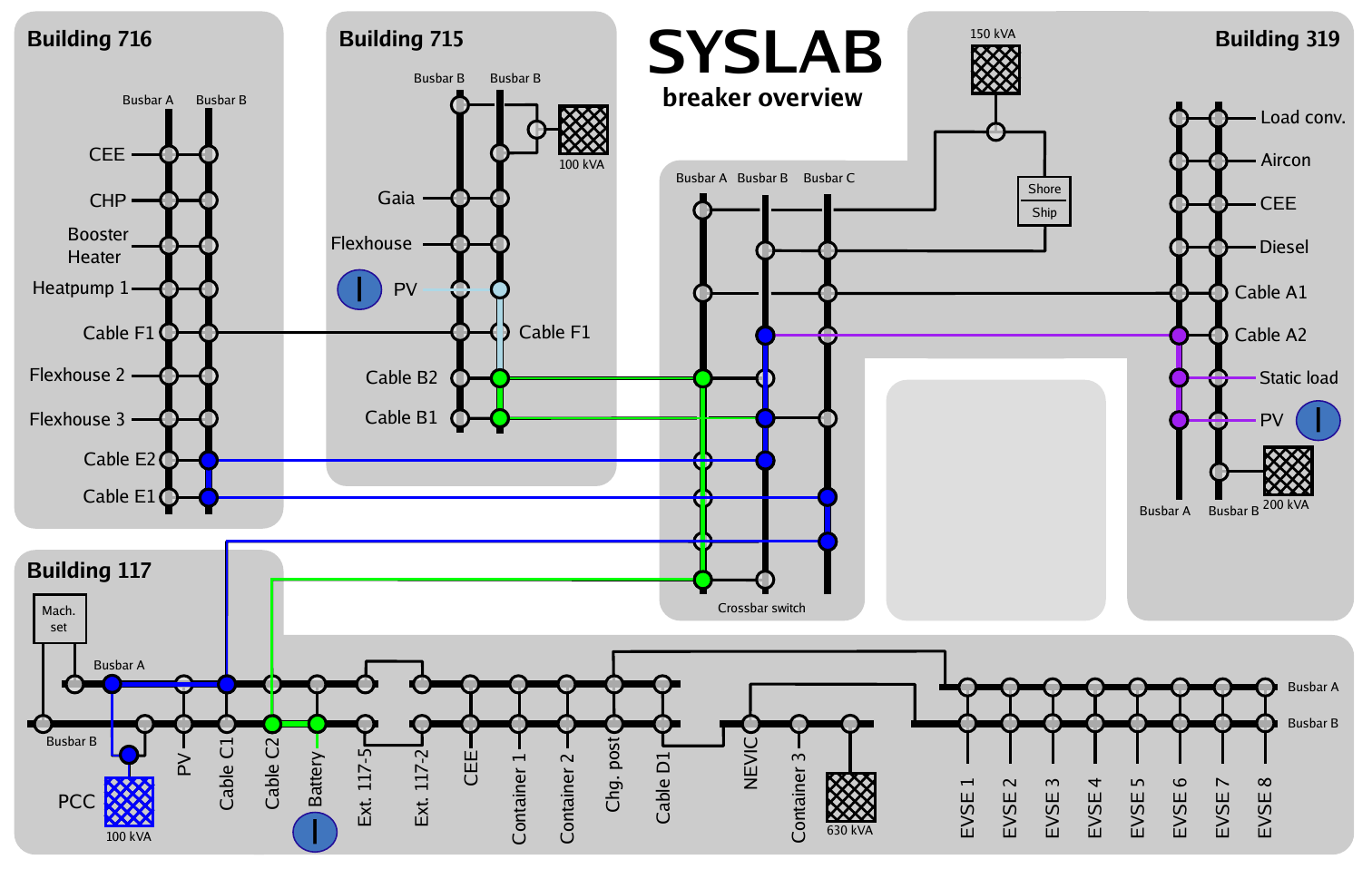}
\caption{SYSLAB infrastructure with the used topology. The colors match the colors in the diagram and in the voltage profile in Figure~\ref{fig:voltage_profile}. Inverters participating in the control algorithm are marked with an I.}
\label{fig:syslab_topology}
\end{figure}

The topology and operational set-points are designed to produce a voltage drop at the beginning and an overvoltage at the end of the feeder. 
Without proper reactive power control, the feeder's ability to host renewable energy infeed is limited and generation would need to be curtailed. The setup consists of the flow battery, two photovoltaic arrays (PV), an adjustable resistive load, and a utility grid connection (PCC). 
This test system is illustrated in  Figure~\ref{fig:voltage_profile}, and Figure~\ref{fig:syslab_topology} presents the corresponding implementation on the SYSLAB topology view.

The active power injection $p_3$ of the battery is interpreted as a renewable source, which is not to be curtailed; its active power infeed is set to $p_3=10$~kW.
The static load is set to an active power consumption of 15 kW ($p_1=-15$~kW) which is larger than the local production, therefore causing a positive active power flow from the substation.
PVs are fluctuating power sources. Therefore, to facilitate repeatability of the experiments and to allow for a comparison between different controllers, the PVs are curtailed to not inject active power ($p_2=$~0~kW).
The different nodes are connected via cables with non-negligible resistance, see Table \ref{tab:cable}. 
Due to a weak link (resistive) cable connecting the battery to the grid, the battery encounters an overvoltage when the reactive power injection is zero.
Both PVs and the battery can measure their voltage magnitudes, and their reactive power injections can be controlled.
The PV inverters have a reactive power range of $\pm$6~kVAr and the battery can be actuated with $\pm8$~kVAr.
The PVs and the battery can communicate with their neighbors, while the load is uncontrolled and unmeasured.
The voltage limits are defined to be 0.95~p.u. and 1.05~p.u.

\begin{table}
\definecolor{bluecable}{RGB}{24, 24, 254}
\definecolor{greencable}{RGB}{24, 254, 24}
\definecolor{purplecable}{RGB}{168, 53, 241}
\definecolor{cyancable}{RGB}{180, 219, 232}
\centering
\caption{Parameters of the cables between busbars/devices.}
\label{tab:cable}
\begin{tabular}{cccccc}
\toprule
    &Cable & Length & Cross section & R  & X \\
    && [m] & [mm\textsuperscript{2}] & [$\Omega$] & [$\Omega$] \\
    \midrule
     \raisebox{1pt}{\textcolor{bluecable}{\rule{4mm}{1mm}}} & C1 & 700 & 240 & 0.085 & 0.054 \\
     \raisebox{1pt}{\textcolor{bluecable}{\rule{4mm}{1mm}}} & E1 & 450 & 240 & 0.055 & 0.035\\
     \raisebox{1pt}{\textcolor{bluecable}{\rule{4mm}{1mm}}} & E2 & 450 & 240 & 0.055 & 0.035\\
     \raisebox{1pt}{\textcolor{purplecable}{\rule{4mm}{1mm}}} & A2 & 25 & 95 & 0.0078 & 0.002\\
     \raisebox{1pt}{\textcolor{purplecable}{\rule{4mm}{1mm}}} & PV1 & 83 & 16 & 0.095 & 0.007\\
     \raisebox{1pt}{\textcolor{purplecable}{\rule{4mm}{1mm}}} & Static Load & 11 & 95 & 0.002 & 0.001\\
     \raisebox{1pt}{\textcolor{greencable}{\rule{4mm}{1mm}}} & B1 & 350 & 95 & 0.11 & 0.027 \\
     \raisebox{1pt}{\textcolor{cyancable}{\rule{4mm}{1mm}}} & PV2 & 8 & 6 & 0.025 & 0.0008\\     
     \raisebox{1pt}{\textcolor{greencable}{\rule{4mm}{1mm}}} & B2 & 350 & 95 & 0.11 & 0.027 \\
     \raisebox{1pt}{\textcolor{greencable}{\rule{4mm}{1mm}}} & C2 & 700 & 240 & 0.085 & 0.054 \\
     \raisebox{1pt}{\textcolor{greencable}{\rule{4mm}{1mm}}} & Battery & 100 & 2.5 & 0.774 & 0.012\\
\bottomrule
\end{tabular}
\end{table}

\section{Experimental Results}
\label{sec:experimentalresults}

\textcolor{highlight}{
In this section, we first demonstrate the suboptimal performance of a decentralized (purely local) controller on the proposed system, by implementing the droop control recommended by the IEEE standard \cite{IEEE1547}. 
We then execute the proposed distributed controller, evaluate its control performance, examine the nested execution of Algorithm~2, hint at a windup phenomenon in case of problem infeasibility, and analyze the trade-off between control performance and communication complexity.
}

{\color{highlight}
\subsection{Local Control}

The droop controller that we implement is the one proposed in \cite{IEEE1547} and similar to the ones suggested in the recent grid codes \cite{VDE,ENTSOE}.
Every inverter implements the following piecewise linear control law.

\vspace{3mm}
\noindent%
\begin{minipage}{0.6\columnwidth}
$\displaystyle
q_h \!=\!
    \begin{cases}
    q_\text{max}& v_h\! < \!v_1\\
    q_\text{max}\dfrac{v_2\!-\!v_h}{v_2\!-v_1} & v_1 \!\leq\! v_h \!\leq\! v_2\\
    0 & v_2\leq v_h\! \leq\! v_3\\
    q_\text{min}\dfrac{v_h\!-\!v_3}{v_4\!-\!v_3} & v_3 \! \leq \! v_h \!\leq\! v_4\\
    q_\text{min}& v_4\! < \!v_h\\
    \end{cases}
$
\end{minipage}%
\begin{minipage}{0.4\columnwidth}
\includegraphics[width=\columnwidth]{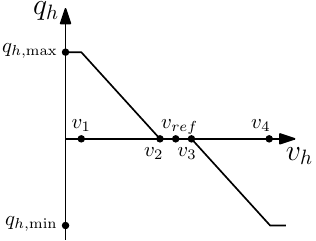}
\end{minipage}%
\vspace{3mm}
where $v_h$ is the measured voltage magnitude, $q_h$ is the calculated reactive power injection, $q_\text{min}$ and $q_\textrm{max}$ are the minimum and maximum reactive power injection. 
We tune the droop curve to $v_1=0.95$~p.u., $v_2=0.99$~p.u., $v_3=1.01$~p.u. and $v_4=1.05$~p.u..

The resulting performance of the controller is reported in Figure~\ref{fig:droop_normal}. When the control is activates at minute 3, only the controller at the battery detects a voltage violation and immediately lowers its reactive power injection to the minimum. However, this is not sufficient to regulate the voltage to the desired voltage range. The PV systems do not detect an overvoltage and therefore do not draw reactive power.
Without introducing coordination between the inverters, the persistent overvoltage at the battery cannot be prevented. Therefore, all local control strategies fail in this setup, as established from a theoretical perspective in \cite{Bolognani2019}.

Figure~\ref{fig:droop_normal} also shows that PV1 injects reactive power around minute 4 of the experiment. This worsens the overvoltage at the battery, which shows that local control decisions can in some cases be even detrimental. 

\begin{figure}[tb]
    \centering
    \begin{footnotesize}
    \input{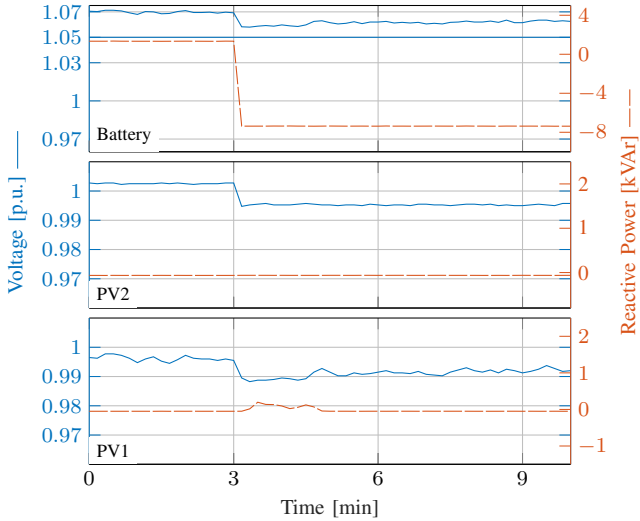}
    \caption{Performance of local droop control (IEEE 1547 standard).}
    \label{fig:droop_normal}
    \end{footnotesize}
\end{figure}

}

\subsection{Controller Evaluation}
\begin{figure}[tb]
    \centering
	\begin{footnotesize}
    \input{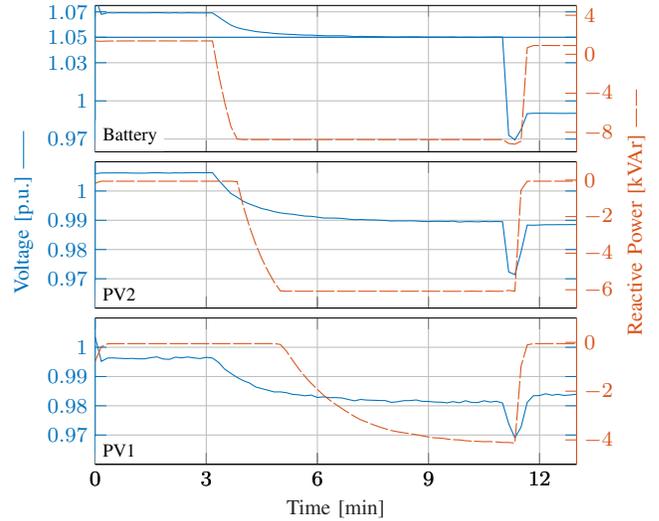}
    \caption{Performance of the distributed voltage controller,  $\alpha=100$.}
    \label{fig:fast_distributed}
    \end{footnotesize}
\end{figure}

Figure~\ref{fig:fast_distributed} shows the performance of the distributed voltage controller with a gain of $\alpha=100$, $K=100$ communication steps to distributively solve the QP, an ascent step length of $\gamma=0.005$ and with matrices $M=X$ and $G=X^{-1}$:
\begin{equation*}
    X\!=\!
    \begin{bmatrix}
    0.10 & 0.09 & 0.09\\
    0.09 & 0.11 & 0.11\\
    0.09 & 0.11 & 0.16
    \end{bmatrix}\!,
    \;
    G\!=\!
    \begin{bmatrix}
    48.3 & -40.7 & 0 \\
    -40.7 & 61.8 & -18.7\\
    0 & -18.7 & 19.1
    \end{bmatrix}\!.
\end{equation*}
\textcolor{highlight}{Cable data have been used to compute the matrix $G$, although the necessary parameters could also be estimated (see \cite{Prostejovsky2016} for an experimental demostration on the same network).}
Notice that, as expected, the matrix $G$ has the sparsity pattern induced by the topology of the distribution grid (zero elements in the positions corresponding to non-neighbors).
The system is initialized with zero reactive power flow.\footnotemark~The controller is activated after 3 minutes and drives all voltages to the desired range. After 11 minutes the active power of the battery, which produces the overvoltage, is brought to 0~kW. The algorithm promptly responds by bringing the reactive power injections of all the power inverters to 0~kVAr.

\footnotetext{Due to an inaccuracy of the sensor used by the internal reactive power controller of the battery, we can observe a small tracking error. The reported measurements in the figures are from accurate sensors.}

\begin{figure}[tb]
    \centering
	\begin{footnotesize}
    \input{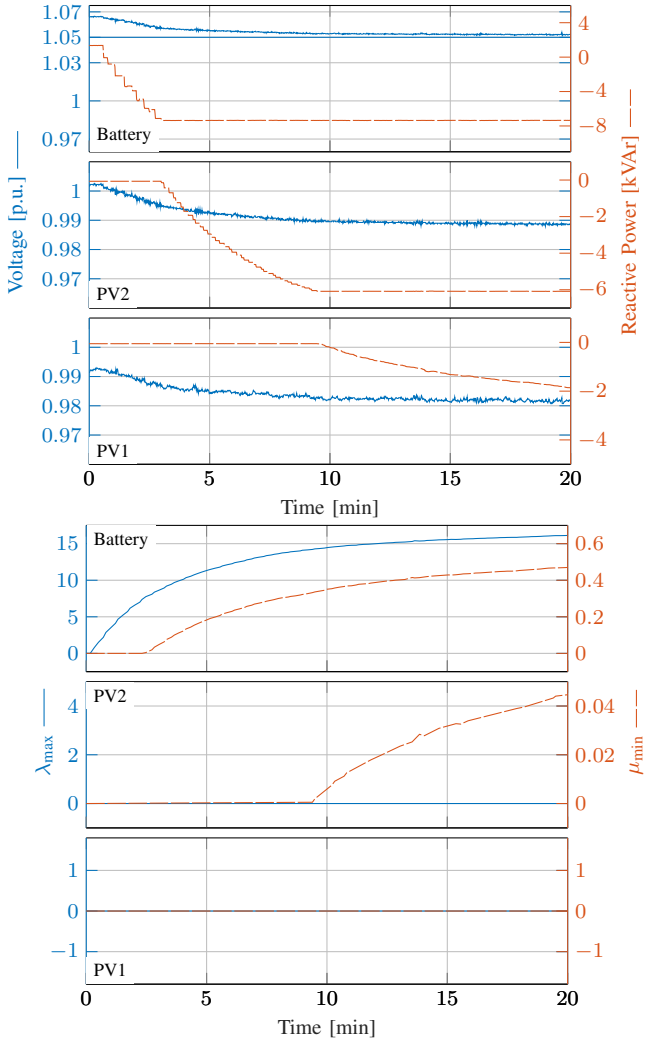}
    \end{footnotesize}
	\begin{footnotesize}
%
%
\definecolor{mycolor1}{rgb}{0.00000,0.44700,0.74100}%
\definecolor{mycolor2}{rgb}{0.85000,0.32500,0.09800}%
\begin{tikzpicture}

\begin{axis}[%
width=\fwidth,
height=0.278\fheight,
at={(0\fwidth,0.594\fheight)},
scale only axis,
xmin=0.00000,
xmax=20.00000,
xticklabels={\empty},
separate axis lines,
every outer y axis line/.append style={mycolor1},
every y tick label/.append style={font=\color{mycolor1}},
every y tick/.append style={mycolor1},
ymin=-2.500000,
ymax=17.50000,
ytick={ 0,  5, 10, 15, 20},
axis background/.style={fill=white},
xmajorgrids,
ymajorgrids
]
\addplot [color=mycolor1, forget plot]
  table[row sep=crcr]{%
0.00000	0.00000\\
0.16667	0.00000\\
0.33333	0.78750\\
0.50696	1.51250\\
0.70914	2.18750\\
0.81547	2.86250\\
1.21357	4.11250\\
1.31529	4.66250\\
1.51666	5.21250\\
1.71818	5.72500\\
1.81899	6.17500\\
2.02266	6.61250\\
2.22480	7.00000\\
2.31740	7.37500\\
2.47932	7.76250\\
2.72895	8.08750\\
2.83187	8.37500\\
3.03199	8.65000\\
3.23349	8.93750\\
3.33461	9.20000\\
3.53675	9.45000\\
3.65839	9.68750\\
3.84065	9.91250\\
4.04284	10.13750\\
4.14389	10.35000\\
4.34451	10.55000\\
4.48648	10.75000\\
4.65372	10.95000\\
4.81726	11.13750\\
4.98505	11.31250\\
5.14887	11.47500\\
5.35565	11.65000\\
5.55779	11.80000\\
5.65887	11.96250\\
5.86100	12.12500\\
6.06328	12.26250\\
6.16418	12.40000\\
6.56839	12.65000\\
6.66946	12.78750\\
6.87703	12.91250\\
6.98301	13.03750\\
7.17494	13.15000\\
7.37712	13.25000\\
7.48427	13.36250\\
7.67831	13.45000\\
7.81368	13.55000\\
7.98130	13.62500\\
8.14514	13.73750\\
8.38691	13.81250\\
8.48816	13.88750\\
8.89042	14.01250\\
8.97759	14.10000\\
9.19447	14.15000\\
9.39596	14.21250\\
9.49676	14.28750\\
9.69985	14.33750\\
9.90159	14.40000\\
10.00280	14.46250\\
10.31902	14.57500\\
10.50867	14.62500\\
10.71084	14.68750\\
10.81763	14.73750\\
11.01425	14.77500\\
11.21637	14.78750\\
11.31753	14.85000\\
11.51969	14.90000\\
11.72211	14.92500\\
11.98692	14.98750\\
12.47924	15.10000\\
12.73236	15.13750\\
12.83311	15.17500\\
13.23611	15.22500\\
13.53867	15.30000\\
13.64343	15.40000\\
13.84498	15.36250\\
14.04622	15.38750\\
14.14698	15.43750\\
14.34831	15.46250\\
14.55003	15.50000\\
14.65110	15.52500\\
14.85327	15.53750\\
15.05576	15.56250\\
15.15679	15.58750\\
15.35893	15.57500\\
15.56106	15.60000\\
15.66222	15.63750\\
15.86426	15.66250\\
16.06642	15.67500\\
16.16745	15.70000\\
16.36961	15.72500\\
16.56538	15.73750\\
16.66654	15.76250\\
16.86787	15.77500\\
17.06941	15.80000\\
17.17701	15.82500\\
17.37923	15.83750\\
17.48027	15.85000\\
17.68242	15.86250\\
17.88452	15.86250\\
17.98559	15.88750\\
18.38844	15.93750\\
18.48923	15.96250\\
18.69109	15.97500\\
18.88332	16.00000\\
18.99738	16.02500\\
19.19820	16.05000\\
19.40028	16.08750\\
19.50133	16.10000\\
19.70363	16.10000\\
19.89055	16.11250\\
20.00791	16.12500\\
};
\end{axis}

\begin{axis}[%
width=\fwidth,
height=0.278\fheight,
at={(0\fwidth,0.594\fheight)},
scale only axis,
xmin=0.00000,
xmax=20.00000,
xticklabels={\empty},
every outer y axis line/.append style={mycolor2},
every y tick label/.append style={font=\color{mycolor2}},
every y tick/.append style={mycolor2},
ymin=-0.10000,
ymax=0.70000,
ytick={  0, 0.2, 0.4, 0.6, 0.8},
axis x line*=bottom,
axis y line*=right
]
\addplot [color=mycolor2, forget plot,dash pattern={on 7pt off 1pt}]
  table[row sep=crcr]{%
0.00000	0.00000\\
2.31740	0.00000\\
2.47932	0.00455\\
2.72895	0.01953\\
2.83187	0.03393\\
3.03199	0.04897\\
3.23349	0.06272\\
3.33461	0.07580\\
3.53675	0.08824\\
3.65839	0.10001\\
3.84065	0.11179\\
4.04284	0.12291\\
4.14389	0.13338\\
4.34451	0.14385\\
4.48648	0.15432\\
4.65372	0.16413\\
4.81726	0.17329\\
4.98505	0.18180\\
5.14887	0.19096\\
5.35565	0.19881\\
5.55779	0.20731\\
5.65887	0.21582\\
6.06328	0.23022\\
6.16418	0.23676\\
6.36627	0.24330\\
6.56839	0.25049\\
6.66946	0.25704\\
6.87703	0.26358\\
6.98301	0.26947\\
7.17494	0.27471\\
7.37712	0.28059\\
7.48427	0.28518\\
7.67831	0.29041\\
7.81368	0.29434\\
7.98130	0.30022\\
8.14514	0.30415\\
8.38691	0.30808\\
8.48816	0.31135\\
8.69032	0.31462\\
8.89042	0.31920\\
8.97759	0.32214\\
9.19447	0.32667\\
9.39596	0.33220\\
9.49676	0.33601\\
9.90159	0.34526\\
10.00280	0.35079\\
10.20558	0.35371\\
10.31902	0.35739\\
10.50867	0.36199\\
10.71084	0.36576\\
10.81763	0.36859\\
11.01425	0.36962\\
11.21637	0.37409\\
11.31753	0.37784\\
11.72211	0.38166\\
11.82309	0.38440\\
12.47924	0.39556\\
12.73236	0.39834\\
12.83311	0.40025\\
13.23611	0.40397\\
13.53867	0.41485\\
13.64343	0.41255\\
13.84498	0.41421\\
14.14698	0.41977\\
14.34831	0.42252\\
15.05576	0.42907\\
15.15679	0.42828\\
15.35893	0.43001\\
15.56106	0.43274\\
15.66222	0.43465\\
15.86426	0.43562\\
16.06642	0.43744\\
16.16745	0.43930\\
16.36961	0.44027\\
17.17701	0.44771\\
17.37923	0.44864\\
17.48027	0.44957\\
17.68242	0.44962\\
17.88452	0.45139\\
17.98559	0.45324\\
18.88332	0.46161\\
18.99738	0.46347\\
19.19820	0.46621\\
19.40028	0.46723\\
19.50133	0.46728\\
19.89055	0.46909\\
20.00791	0.47002\\
};
\node[anchor=north west, fill=white] at (current axis.north west) {\scriptsize Battery};
\end{axis}

\begin{axis}[%
width=\fwidth,
height=0.278\fheight,
at={(0\fwidth,0\fheight)},
scale only axis,
xmin=0.00000,
xmax=20.00000,
xlabel style={font=\color{white!15!black}},
xlabel={Time [min]},
separate axis lines,
every outer y axis line/.append style={mycolor1},
every y tick label/.append style={font=\color{mycolor1}},
every y tick/.append style={mycolor1},
ymin=-1.80000,
ymax=1.80000,
ytick={  -1, 0, 1},
axis background/.style={fill=white},
xmajorgrids,
ymajorgrids
]
\addplot [color=mycolor1, forget plot]
  table[row sep=crcr]{%
0.00000	0.00000\\
20.00791	0.00000\\
};
\end{axis}

\begin{axis}[%
width=\fwidth,
height=0.278\fheight,
at={(0\fwidth,0\fheight)},
scale only axis,
xmin=0.00000,
xmax=20.00000,
every outer y axis line/.append style={mycolor2},
every y tick label/.append style={font=\color{mycolor2}},
every y tick/.append style={mycolor2},
ymin=-1.80000,
ymax=1.8000,
ytick={  -1, 0,  1},
axis x line*=bottom,
axis y line*=right
]
\addplot [color=mycolor2, forget plot,dash pattern={on 7pt off 1pt}]
  table[row sep=crcr]{%
0.00000	0.00000\\
20.00791	0.00000\\
};
\node[anchor=south west, fill=white] at (current axis.south west) {\scriptsize PV1};
\end{axis}

\begin{axis}[%
width=\fwidth,
height=0.278\fheight,
at={(0\fwidth,0.297\fheight)},
scale only axis,
xmin=0.00000,
xmax=20.00000,
xticklabels={\empty},
separate axis lines,
every outer y axis line/.append style={mycolor1},
every y tick label/.append style={font=\color{mycolor1}},
every y tick/.append style={mycolor1},
ymin=-1.0000,
ymax=5.0000,
ytick={  0, 2, 4},
ylabel style={font=\color{mycolor1}},
ylabel={$\lambda_\textrm{max}$},
ylabel style={alias=auy_v},
axis background/.style={fill=white},
xmajorgrids,
ymajorgrids
]
\addplot [color=mycolor1, forget plot]
  table[row sep=crcr]{%
0.00000	0.00000\\
20.00759	0.00000\\
};
\end{axis}
\draw[mycolor1] (auy_v.east) -- ++(0,0.5);

\begin{axis}[%
width=\fwidth,
height=0.278\fheight,
at={(0\fwidth,0.297\fheight)},
scale only axis,
xmin=0.00000,
xmax=20.00000,
xticklabels={\empty},
every outer y axis line/.append style={mycolor2},
every y tick label/.append style={font=\color{mycolor2}},
every y tick/.append style={mycolor2},
y tick label style={
        /pgf/number format/.cd,
            fixed,
            scaled ticks=false,
            precision=3,
        /tikz/.cd
    },
ymin=-0.01,
ymax=0.05,
ytick={ 0, 0.02,0.04},
ylabel style={font=\color{mycolor2}},
ylabel={$\mu_\textrm{min}$},
ylabel style={alias=auy_q},
axis x line*=bottom,
axis y line*=right
]
\addplot [color=mycolor2, forget plot,dash pattern={on 7pt off 1pt}]
  table[row sep=crcr]{%
0.00000	0.00000\\
9.39554	0.00052\\
9.49632	0.00193\\
10.20526	0.00755\\
10.32365	0.00915\\
10.71068	0.01120\\
10.81742	0.01257\\
11.82274	0.01790\\
12.32039	0.02017\\
13.03510	0.02352\\
13.53835	0.02528\\
13.84452	0.02839\\
14.04577	0.02782\\
14.65152	0.03084\\
15.35862	0.03275\\
15.56074	0.03258\\
15.86382	0.03394\\
17.88496	0.03898\\
18.18794	0.03958\\
19.40075	0.04327\\
19.50111	0.04405\\
20.00759	0.04464\\
};
\node[anchor=north west, fill=white] at (current axis.north west) {\scriptsize PV2};
\end{axis}
\draw[dash pattern={on 7pt off 1pt},mycolor2] (auy_q.east) -- ++(0,0.5);
\node at (0,0) [xshift=-\fmargin] {};
\node at (\fwidth,0) [xshift=\fmargin] {};
\end{tikzpicture}%
    \caption{Behavior of the distributed controller with control gain $\alpha=50$. Upper panels: electrical quantities $v, q$. Lower panel: dual multipliers $\lambda_\textrm{max}$ and $\mu_\textrm{min}$ (evaluated at the last step of Algorithm~2).}
    \label{fig:multipliers}
    \end{footnotesize}
\end{figure}

For a more in-depth analysis of the control behavior we provide the data in Figure~\ref{fig:multipliers} for a controller with $\alpha=50$ and $K=50$.
We report both the electrical quantities $v$ and $q$ and the controllers' internal variables $\lambda_\text{max}$ and $\mu_\text{min}$ ($\lambda_\text{min}$ and $\mu_\text{max}$ remain zero in this experiment).
Once the controller is activated at 3 minutes, the voltage violation leads to a growing $\lambda_{3,\text{max}}$ at the battery. 
As this integral variable grows, the battery starts drawing reactive power. Once the reactive power $q_3$ of the battery reaches the battery's reactive power limit, the corresponding multiplier $\mu_{3,\textrm{min}}$ starts growing. At each iteration of Algorithm~2, this value is communicated to PV2. Ultimately, PV2 starts drawing reactive power as well (thus participating to the voltage regulation task). Once the reactive power limit of PV2 is reached, its $\mu_{2,\textrm{min}}$ value becomes positive and PV1 starts to draw reactive power.
As long as there remains an overvoltage at the battery, the battery keeps integrating its $\lambda_{3,\textrm{max}}$, which leads to a larger reactive power demand by the inverter that is closer to battery and is not yet saturated. Finally, the voltage converges to the voltage constraint. Once that point is reached the system has settled (not fully represented in Figure~\ref{fig:multipliers}).
Three remarks are due.
\begin{itemize}
    \item There is no central clock signal and the different inverters time their iterations of Algorithm~1 independently. Measurements are therefore not perfectly synchronous. We do not observe any detrimental effect in the experiment.
    \item Each controller gathers raw voltage measurements. No filtering or state estimation is performed (which, in general, would require a system model and further exchange of information). The control performance seems to be unaffected by measurement noise and quantization.
    \item With a smaller actuation interval (smaller T) and therefore more frequent control actuations the settling time of the algorithm can be reduced.
\end{itemize}

\subsection{Convergence of Algorithm 2}

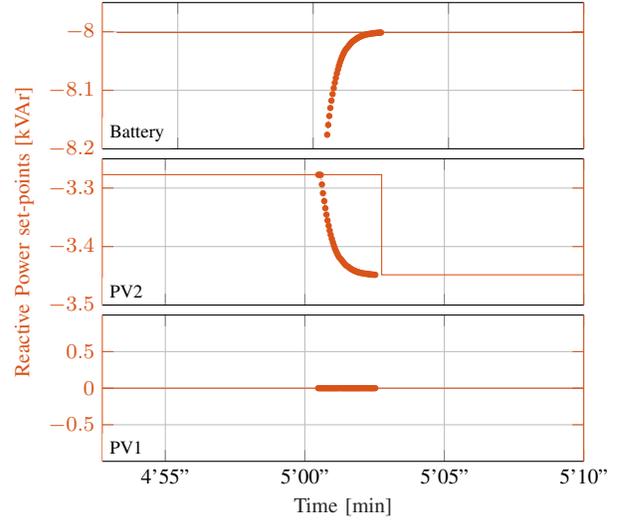
\begin{figure}[tb]
    \centering
	\begin{footnotesize}
%
%
\definecolor{mycoloralt1}{rgb}{0.85000,0.32500,0.09800}
\definecolor{mycoloralt2}{rgb}{0.85000,0.32500,0.09800}
\begin{tikzpicture}

\begin{axis}[%
width=\fwidth,
height=0.278\fheight,
at={(0\fwidth,0.594\fheight)},
scale only axis,
xmin=4.9202,
xmax=5.2166,
xtick={4.96666,5.05,5.13333,5.2166},
xticklabels={\empty},
separate axis lines,
every outer y axis line/.append style={mycoloralt1},
every y tick label/.append style={font=\color{mycoloralt1}},
every y tick/.append style={mycoloralt1},
ymin=-8.20000,
ymax=-7.95000,
axis background/.style={fill=white},
xmajorgrids,
ymajorgrids
]
\addplot [color=mycoloralt1, forget plot]
  table[row sep=crcr]{%
4.92902	-8.00119\\
5.09567	-8.00119\\
5.09568	-8.00128\\
5.26235	-8.00128\\
};
\addplot [color=mycoloralt2, dotted, mark=*, mark options={solid, mycoloralt2},  mark size=1, forget plot]
  table[row sep=crcr]{%
5.05866	-8.17598\\
5.05928	-8.15916\\
5.05996	-8.14396\\
5.06059	-8.13020\\
5.06127	-8.11776\\
5.06199	-8.10651\\
5.06269	-8.09633\\
5.06334	-8.08713\\
5.06397	-8.07880\\
5.06461	-8.07128\\
5.06525	-8.06447\\
5.06587	-8.05831\\
5.06648	-8.05274\\
5.06711	-8.04770\\
5.06774	-8.04314\\
5.06840	-8.03902\\
5.06902	-8.03529\\
5.06967	-8.03192\\
5.07049	-8.02887\\
5.07136	-8.02611\\
5.07206	-8.02362\\
5.07266	-8.02136\\
5.07329	-8.01932\\
5.07394	-8.01747\\
5.07467	-8.01580\\
5.07550	-8.01429\\
5.07614	-8.01293\\
5.07678	-8.01169\\
5.07741	-8.01058\\
5.07803	-8.00956\\
5.07870	-8.00865\\
5.07933	-8.00782\\
5.07996	-8.00708\\
5.08059	-8.00640\\
5.08123	-8.00579\\
5.08186	-8.00524\\
5.08256	-8.00474\\
5.08340	-8.00428\\
5.08426	-8.00387\\
5.08494	-8.00350\\
5.08558	-8.00317\\
5.08636	-8.00287\\
5.08722	-8.00259\\
5.08785	-8.00234\\
5.08850	-8.00212\\
5.08919	-8.00192\\
5.08982	-8.00173\\
5.09045	-8.00157\\
5.09115	-8.00142\\
5.09185	-8.00128\\
};
\node[anchor=south west, fill=white] at (current axis.south west) {\scriptsize Battery};
\end{axis}

\begin{axis}[%
width=\fwidth,
height=0.278\fheight,
at={(0\fwidth,0.297\fheight)},
scale only axis,
xmin=4.92902,
xmax=5.2166,
xtick={4.96666,5.05,5.13333,5.2166},
xticklabels={\empty},
separate axis lines,
every outer y axis line/.append style={mycoloralt1},
every y tick label/.append style={font=\color{mycoloralt1}},
every y tick/.append style={mycoloralt1},
ymin=-3.5000,
ymax=-3.25000,
ylabel style={font=\color{mycoloralt1}},
ylabel={Reactive Power set-points [kVAr]},
axis background/.style={fill=white},
xmajorgrids,
ymajorgrids
]
\addplot [color=mycoloralt1, forget plot]
  table[row sep=crcr]{%
4.92926	-3.27715\\
5.09591	-3.27715\\
5.09592	-3.44831\\
5.26259	-3.44831\\
};
\addplot [color=mycoloralt2, dotted, mark=*, mark options={solid, mycoloralt2},  mark size=1, forget plot]
  table[row sep=crcr]{%
5.05804	-3.27727\\
5.05886	-3.27738\\
5.05951	-3.27748\\
5.06020	-3.29394\\
5.06082	-3.30882\\
5.06160	-3.32228\\
5.06224	-3.33445\\
5.06291	-3.34547\\
5.06357	-3.35542\\
5.06424	-3.36443\\
5.06485	-3.37258\\
5.06545	-3.37995\\
5.06611	-3.38661\\
5.06674	-3.39264\\
5.06734	-3.39809\\
5.06794	-3.40302\\
5.06858	-3.40748\\
5.06925	-3.41152\\
5.06986	-3.41516\\
5.07054	-3.41846\\
5.07153	-3.42145\\
5.07224	-3.42415\\
5.07293	-3.42659\\
5.07354	-3.42880\\
5.07418	-3.43079\\
5.07483	-3.43260\\
5.07578	-3.43423\\
5.07648	-3.43571\\
5.07701	-3.43705\\
5.07763	-3.43826\\
5.07828	-3.43935\\
5.07894	-3.44034\\
5.07958	-3.44123\\
5.08022	-3.44204\\
5.08084	-3.44277\\
5.08145	-3.44344\\
5.08211	-3.44403\\
5.08273	-3.44458\\
5.08342	-3.44507\\
5.08443	-3.44551\\
5.08514	-3.44591\\
5.08582	-3.44627\\
5.08645	-3.44660\\
5.08731	-3.44689\\
5.08808	-3.44716\\
5.08874	-3.44741\\
5.08939	-3.44762\\
5.09005	-3.44782\\
5.09068	-3.44800\\
5.09131	-3.44816\\
5.09209	-3.44831\\
};
\node[anchor=south west, fill=white] at (current axis.south west) {\scriptsize PV2};
\end{axis}

\begin{axis}[%
width=\fwidth,
height=0.278\fheight,
at={(0\fwidth,0\fheight)},
scale only axis,
xmin=4.92902,
xmax=5.2166,
xlabel style={font=\color{white!15!black}},
xlabel={Time [min]},
xtick={4.96666,5.05,5.13333,5.2166},
xticklabels={4'55'',5'00'',5'05'',5'10''},
ytick={-0.5,0,0.5},
separate axis lines,
every outer y axis line/.append style={mycoloralt1},
every y tick label/.append style={font=\color{mycoloralt1}},
every y tick/.append style={mycoloralt1},
ymin=-1.00000,
ymax=1.00000,
axis background/.style={fill=white},
xmajorgrids,
ymajorgrids
]
\addplot [color=mycoloralt1, forget plot]
  table[row sep=crcr]{%
4.92927	0.00000\\
5.26261	0.00000\\
};
\addplot [color=mycoloralt2, dotted, mark=*, mark options={solid, mycoloralt2}, mark size=1, forget plot]
  table[row sep=crcr]{%
5.05801	0.00000\\
5.05886	0.00000\\
5.05949	0.00000\\
5.06020	0.00000\\
5.06084	0.00000\\
5.06148	0.00000\\
5.06216	0.00000\\
5.06287	0.00000\\
5.06355	0.00000\\
5.06420	0.00000\\
5.06482	0.00000\\
5.06549	0.00000\\
5.06609	0.00000\\
5.06671	0.00000\\
5.06732	0.00000\\
5.06795	0.00000\\
5.06858	0.00000\\
5.06919	0.00000\\
5.06987	0.00000\\
5.07061	0.00000\\
5.07142	0.00000\\
5.07220	0.00000\\
5.07289	0.00000\\
5.07349	0.00000\\
5.07417	0.00000\\
5.07490	0.00000\\
5.07561	0.00000\\
5.07634	0.00000\\
5.07699	0.00000\\
5.07760	0.00000\\
5.07828	0.00000\\
5.07893	0.00000\\
5.07956	0.00000\\
5.08019	0.00000\\
5.08081	0.00000\\
5.08146	0.00000\\
5.08207	0.00000\\
5.08275	0.00000\\
5.08344	0.00000\\
5.08434	0.00000\\
5.08512	0.00000\\
5.08579	0.00000\\
5.08656	0.00000\\
5.08733	0.00000\\
5.08811	0.00000\\
5.08872	0.00000\\
5.08943	0.00000\\
5.09003	0.00000\\
5.09064	0.00000\\
5.09128	0.00000\\
5.09211	0.00000\\
};
\node[anchor=south west, fill=white] at (current axis.south west) {\scriptsize PV1};
\end{axis}
\node at (0,0) [xshift=-\fmargin] {};
\node at (\fwidth,0) [xshift=\fmargin] {};
\end{tikzpicture}%
    \caption{Convergence of one execution of Algorithm 2. Reactive power set-points (solid) and updates of the internal variable $\hat q$ (dotted).}
    \label{fig:q_and_q_head}
    \end{footnotesize}
\end{figure}

In Figure~\ref{fig:q_and_q_head}, we can see how the internal variables $\hat q$ of the three inverters converge during the execution of Algorithm~2. The algorithm is started as soon as the multipliers $\lambda_\text{max}$ are updated with the measured voltage violation. 
Agents update their internal variable $\hat {q}_h$ (orange dots in Figure~\ref{fig:q_and_q_head}) and their multipliers $\mu_{h,\text{min}}, \mu_{h,\text{max}}$ (not represented) while communicating with their neighbors at each iteration.
After $K$ iterations, the internal value $\hat q$ is used to actuate the system by updating the reactive power set-points for the inverters (blue line).
A few remarks are due:
\begin{itemize}
    \item due to the warm start of the algorithm and the relatively small changes in $\lambda_\textrm{max}$, the initialization of $\hat q$ is already close to the final (optimal) value;
    \item \textcolor{highlight}{$K=40$ iterations suffice for the convergence of Algorithm~2 in this experiment (see Section~\ref{sec:scaling} for further discussion on the effect of early termination of Algorithm~2);} 
    \item the time needed to complete Algorithm~2 is significantly shorter than the sampling rate of Algorithm~1 (10 s).
\end{itemize}

\subsection{Controller Windup}
\begin{figure}[tb]
    \centering
	\begin{footnotesize}
    \input{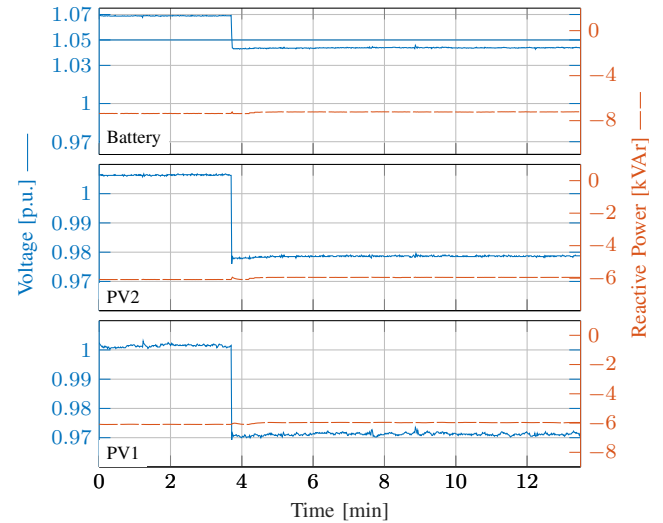}
    \caption{Controller windup due to a persistent overvoltage.}
    \label{fig:distributed_windup}
    \end{footnotesize}
\end{figure}

{\color{highlight}
Figure~\ref{fig:distributed_windup} illustrates the behavior of the proposed scheme when the Volt/VAr regulation problem is temporarily unfeasible. A persistent overvoltage at the battery leads to a constantly growing $\lambda_{3,\text{max}}$. All inverters are drawing their maximum reactive power, which confirms that the voltage cannot be regulated: there does not exist a feasible reactive power input such that all voltages are within the voltage limits.

Once we remove the cause of the overvoltage (at approx. 4 minutes) and the voltage drops, the inverters do not adjust their reactive power injection, but remain saturated at their limit value for several minutes.
This phenomenon corresponds to the windup behavior that is often observed in integral controllers.
Here, the integrator is $\lambda_{3,\text{max}}$ of the battery.

One solution to this windup problem is to stop the integration of the voltage violation once all inverters have saturated. While this is an easy modification for a centralized controller (see \cite{Ortmann2020}), a more sophisticated anti-windup scheme is needed in a distributed setup, where no single agent is aware of the infeasibility of the optimization problem. The design of an effective distributed anti-windup scheme is an interesting and open problem \emph{per se}.
}

\subsection{Control Performance vs Communication Complexity}

The ability of performing optimal voltage control without global communication comes at a price. 
As detailed in Section~\ref{ssec:distributing}, in order to obtain an iterative update that only requires neighbor-to-neighbor communication we had to constrain the choice of the quadratic cost parameter $M$ in \eqref{eq:optimization_problem}.

We showed that $M=X$ is a valid choice, $X$ being the grid susceptance matrix.
As discussed in \cite{Bolognani2015}, the minimization of $q^T X q$ is connected to the minimization of power losses caused by reactive power flows (under the assumption of homogeneous X/R ratio). 
Moreover, as discussed in \cite{Bolognani2019}, the cost $q^T X q$ can then be rewritten as $(Xq)^T G X q$, where $X q$ is the first order approximation of the voltage drop caused by reactive power injection. Therefore, because $G$ has the structure of a Laplacian, $q^T X q$ promotes equal voltage drops in the network.

In general, however, a network operator may be interested in minimizing a different cost function, e.g.
\begin{equation}
J_\text{fair} :=
\sum_i \left(q_i/q_i^\text{max}\right)^2
\label{eq:jfair}
\end{equation}
which promotes proportional fairness in the use of the reactive power capacity of each inverter.
The difference in the reactive power set-points and in the resulting cost is reported in Table~\ref{tab:efficiency}.
Given the inexpensive nature of reactive power, these differences are in most cases acceptable.

\begin{table}
\centering
\caption{Comparison between the steady-state of the distributed algorithm and the maximal-fairness set-points that minimize \eqref{eq:jfair}.}
\label{tab:efficiency}
\begin{tabular}{@{}cccc@{}}
\toprule
     & $\arg\min\ q^TXq$ & $\arg\min\ J_\text{fair}$ & difference \\
    \midrule
     $J_\text{fair}$ & 2.12 & 1.98 & 6.9\% \\
     PV1 $q_1$ & -2.06 kVAr & -3.76 kVAr & 0.28 [p.u.] \\
     PV2 $q_2$ & -6 kVAr & -4.6 kVAr & 0.23 [p.u.] \\
     Battery $q_3$ & -8 kVAr & -8 kVAr & 0 [p.u.] \\
\bottomrule
\end{tabular}
\end{table}

{\color{highlight}
\section{Scalability}
\label{sec:scaling}

In this section we investigate how the performance of the proposed feedback scheme scales with the number of nodes. In order to perform this analysis, we consider a fictitious scenario which is identical to the one described in Section~\ref{ssec:benchmark}, but where $N'$ extra ``dummy'' nodes have been added on the line connecting PV2 to the Battery (see Figure~\ref{fig:additional_nodes}). These nodes are equally spaced and have zero reactive power capability. They can communicate one to the next one, so that the entire communication graph becomes a line of $N = N' + 3$ nodes. Their presence therefore affect the execution of the algorithm without affecting the optimal solution of the problem. 

\begin{figure}[tb]
    \centering
    \includegraphics[scale=0.85]{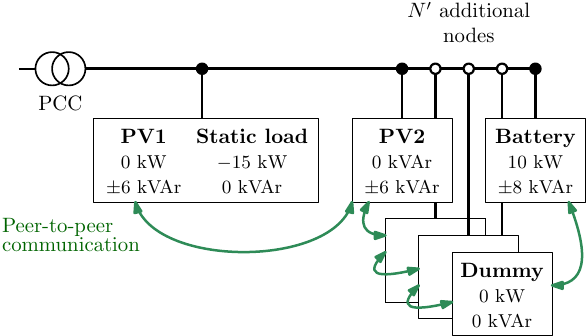}\\[1mm]
    \caption{\color{highlight} Diagram of the electrical topology and of the communication graph used in the numerical analysis of the algorithm scalability.}
    \label{fig:additional_nodes}
\end{figure}

We compare two implementations of our method: in the first case (that we denote as $K\gg 1)$, we allow an arbitrarily large number $K$ of communication steps between each actuation step (namely, we allow communication until convergence up to a tolerance of $10$~VAr); in the second case, we only allow one communication step for each actuation step ($K=1$).

This second case closely resembles what was proposed \cite{Bolognani2015}.
More generally, we consider it as a prototype for the other distributed methods available in the literature, where communication and actuation are always interleaved one-to-one.
These include for example the primal-dual methods proposed in \cite{Magnusson2019} or in \cite{Qu2019}.
The other methods reviewed in the Introduction also share the same interleaving between communication and actuation. We will see in this section how this appears to be a design choice that limits performance as the network grows in size.
In the comparison, it is important to keep in mind that iterations that only require computation and communication can be executed much faster than iterations that require actuation of the system and measurement. We therefore counted and reported them separately, as \emph{communication steps} and \emph{actuation steps}.

We executed the algorithm with $K\gg 1$ and $K=1$ for networks of different sizes, and these are the main findings.
\smallskip

\noindent\textbf{Ease of tuning:}
\begin{itemize}
\item If $K$ is large, then the tuning of the inner optimization gain $\gamma$ becomes very simple; Figure~\ref{fig:tuninggamma} shows how a large $K$ gives a plateau of valid choices for $\gamma$.
\item An optimal value of $\gamma$ as a function of the grid parameters has been suggested in \cite[Corollary 2]{Bolognani2015} when $K$ is unbounded, and seems to be an excellent choice also for $K$ finite but sufficiently large. In the specific case of Figure~\ref{fig:tuninggamma}, the recommended $\gamma$ is $1/(2\sigma(G)) = 0.004$.
\item For large $K$, tuning $\alpha$ becomes significantly simpler. All the executions of $K\gg 1$ in Table \ref{tab:comparison} use the same parameter $\alpha=100$ and attain similar time to convergence, while $\alpha$ needed to be substantially re-tuned when $K=1$.
\end{itemize}

\noindent\textbf{Controller performance:}
\begin{itemize}
\item By allowing many communication steps, the number of actuation steps required for convergence is significantly reduced (see Table \ref{tab:comparison}).
\item The number of communication steps required for convergence of the nested subproblem increases with the size of the network (although not exponentially). Remember that these steps only require communication and computation, and we showed that they may also be performed asynchronously (in contrast to the actuation steps, which need to be synchronous).
By using an asynchronous implementation of the inner loop data exchange, the time needed for a single communication step is only determined by the communication speed between two neighbouring nodes \cite{semitsogloutsiapos2019}.
\item The performance of the controller degrades gracefully if an upper bound on the communication steps is imposed (see Table \ref{tab:truncated}).
\end{itemize}

\begin{figure}[bt]
    \centering
    \begin{footnotesize}
    \definecolor{mycolor1}{rgb}{0.00000,0.44700,0.74100}%
\definecolor{mycolor2}{rgb}{0.85000,0.32500,0.09800}%
\begin{tikzpicture}

\begin{loglogaxis}[%
width=1.1\fwidth,
height=0.5\fheight,
at={(0\fwidth,0\fheight)},
scale only axis,
xmin=0.00008,
xmax=0.025,
xtick={ 0.0001,  0.0002, 0.0005, 0.001, 0.002, 0.005, 0.01, 0.02},
xticklabels={ 0.0001,  0.0002, 0.0005, 0.001, 0.002, 0.005, 0.01, 0.02},
xlabel={Tuning parameter $\gamma$ of Algorithm~2},
ymin=0.00000,
ymax=16000,
ytick={ 1,  10, 100, 1000, 10000},
ylabel style={align=center}, ylabel=Number of actuation\\ steps for convergence,
yminorticks=false,
ymajorgrids
]
\addplot [color=blue, mark=square*]
  table[row sep=crcr]{%
0.0005	8861\\
0.001	3376\\
0.002	1265\\
0.005	346\\
0.01	147\\
0.02	81\\
};

\addplot [color=orange,mark=diamond*,mark size=3pt]
  table[row sep=crcr]{%
0.0001	3213\\
0.0002	1183\\
0.0005	318\\
0.001	141\\
0.002	78\\
0.005	58\\
0.01	64\\
};

\addplot [color=black,mark=triangle*,mark size=3pt]
  table[row sep=crcr]{%
0.0001	141\\
0.0002	78\\
0.0005	58\\
0.001	64\\
0.002	67\\
0.005	67\\
0.01	67\\
};

\legend{K=1,K=10,K=100}
\end{loglogaxis}

\end{tikzpicture}%
    \caption{\color{highlight} Number of actuation steps required for convergence as a function of $\gamma$ and of the number of communication steps $K$. Network with 3 inverters and $\alpha = 100$. Each curve stops, at the right end, at the largest $\gamma$ that does not cause instability.}
    \label{fig:tuninggamma}
    \end{footnotesize}
\end{figure}
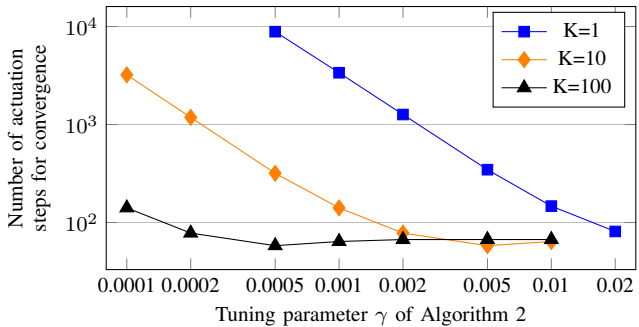

\begin{table}[bt]
    \centering
    \color{highlight}
    \caption{Actuation steps required for convergence. For $K\gg 1$, parameters are constant and are $\alpha=100$, $\gamma=1/(2\sigma(G))$. For $K=1$, $\alpha$ has been optimized for every instance.}
    \label{tab:comparison} 
    \begin{tabular}{@{}cccc@{}}
        \toprule 
        & \multicolumn{2}{c}{$K\gg 1$} & $K=1$\\
         \cmidrule(lr){2-3} 
        nodes & communication steps & actuation steps & actuation steps \\
         \midrule 
         3 & 35 & 46 & 258 ($\alpha=40$) \\
         7 & 566 & 46 & 2975 ($\alpha=3.1$)\\
         10 & 1417 & 47 & 5972 ($\alpha=1.6$)\\
         30 & 14515 & 51 & >30000 \\
         100 & 19848 & 61 & >30000 \\
         \bottomrule
    \end{tabular}
\end{table}

\begin{table}[bt]
    \centering
    \color{highlight}
    \caption{Actuation steps required for convergence, for different values of $K$ (communication steps) in a network of 30 nodes.}
    \label{tab:truncated}  
    \begin{tabular}{@{}lcccccc@{}}
        \toprule 
        $K$ & 100 & 300 & 1000 & 3000 & 10000 & 30000 \\
        actuation steps & 724 & 211 & 67 & 39 & 50 & 51 \\
         \bottomrule
    \end{tabular}
\end{table}

These findings indicate how the decomposition of the iterative optimization scheme into an iteration that requires actuation of the grid (and therefore cannot be executed too frequently) and a nested sequence of communication steps is fundamental for the overall scalability of the solution.
}

\section{Conclusion}
We implemented for the first time a fully distributed peer-to-peer Volt/VAr controller on a real low-voltage distribution network. 
The controller at each inverter only uses local voltage measurements and the required model knowledge is only  the electrical distance to its neighbors. 
No filtering or centralized estimation is needed, and the controller is able to drive the system to an optimal point where all voltage and reactive power constraints are satisfied.
Each inverter is allowed to communicate only with its neighbors in the electric topology.
We also showed that the performance of such a distributed strategy scales nicely with the size of the grid, as long as the communication rate is substantially higher than the measurement/actuation rate.
Moreover, we highlight some directions for future investigation, such as optimizing the trade-off between communication complexity and performance, detecting problem infeasibility, and analyzing finite-time convergence of the nested algorithm.

\section{Acknowledgment}
The authors are grateful to Dimitrios Semitsoglou-Tsiapos for his support in adapting his distributed framework to the control algorithm at hand.

\bibliographystyle{IEEEtran}
\bibliography{IEEEabrv,biblio}

\end{document}